\documentclass{ws-ijmpe}

\begin{document}

\markboth{Yingxun Zhang}{Probing the density dependence of symmetry
energy at subsaturation density with HICs}

\catchline{}{}{}{}{}

\title{Probing the density dependence of symmetry energy at subsaturation density with
HICs}

\author{\footnotesize Yingxun Zhang, M.B.Tsang, Zhuxia Li, P. Danielewicz, W.G. Lynch, Xiaohua Lu}


\address{Department of Nuclear Physics, China Institute of Atomic Energy, \\
Beijing, 102413, China\\
JINA/NSCL, Michigan State University, \\
East Lansing, MI, 48824, USA\\
}

\address{}

\maketitle


\begin{abstract}
The reaction mechanism of the central collisions and peripheral
collisions for $^{112,124}Sn+^{112,124}Sn$ at $E/A=50MeV$ is
investigated  within the framework of the Improved Quantum Molecular
Dynamics model. The results show that multifragmentation process is
an important mechanism at this energy region, and the influence of
the cluster emission on the double n/p ratios and the isospin
transport ratio are important. Furthermore, three observables,
double n/p ratios, isospin diffusion and the rapidity distribution
of the ratio $R_{7}$ for $^{112,124}Sn+^{112,124}Sn$ at E/A=50MeV
are analyzed with the Improved Quantum Molecular Dynamics model. The
results show that these three observables are sensitive to the
density dependence of the symmetry energy. By comparing teh
calculation results to the data, the consistent constraint on the
density dependence of the symmetry energy from these three
observables is obtained.

\end{abstract}

\section{Introduction}

The Equation of State (EOS) of asymmetric nuclear matter can be
written approximately as
$E(\rho,\delta)=E(\rho,\delta=0)+E_{sym}(\rho)\delta^{2}+\mathcal{O}(\delta^4)$
, with $\delta=(\rho_{n}-\rho_{p})/(\rho_{n}+\rho_{p})$. It closely
relates to different areas of nuclear physics, such as the nuclear
structure of finite nuclei, dynamics process of neutron rich heavy
ion collisions, physics of neutron star,
et.al.\cite{BALi2008,Steiner2005,Pawel2003,Pawel2009,Steiner2008,Todd2005,Latt2004}.
For the symmetry nuclear matter, measurements of isoscalar
collective vibrations, collective flow and kaon production in
energetic nucleus-nucleus collisions have constrained the EOS from
normal density to five times saturation
density\cite{Pawel2002,Fuchs2006,Young1999}. However, large
uncertainties exist on the theoretical prediction on the density
dependence of the EOS for neutron matter \cite{Brown2000}. So,
determining the density dependence of the symmetry energy,
$E_{sym}(\rho)$, becomes one of the main goals in nuclear physics at
present and in the near future and has stimulated many theoretical
and experimental studies. Heavy ion collisions with neutron-rich
nuclei provide a unique opportunity to obtain the information of the
density dependence of the symmetry energy in the laboratories
because large extent of density can be formed during the HICs. Many
useful observables from HICs, such as isoscaling of the isotope
\cite{HSXu2000,Tsang2001,Shetty2004}, isospin diffusion
\cite{Tsang2004,LWChen2005,TXLiu2007,BALi2005}, neutron to proton
yield ratios and its flow at intermediate heavy ion collisions have
been proposed and studied both theoretically and experimentally in
order to constrain the density dependence of symmetry energy at
subsaturation density. And $\pi^{-}/\pi^{+}$ ratios and its flow in
high energy HICs was proposed to constrain the density dependence of
the symmetry energy at supersaturation
density\cite{Zhang2008,Fami2006,BALi1997,BALi2006,BALi2000,GCYo2006,Trau2009,ZGXiao2009}.

Since the complex cluster production are observed in experiment for
intermediate energy HICs, it is necessary to investigate the effects
of clusters emission on these observables, neutron to proton ratios
and isospin diffusion. As a microscopic dynamical n-body transport
theory, formation of fragments is included automatically in the QMD
type model. So, the
ImQMD05\cite{Zhang2008,Zhang2005,Zhang2006,Zhang2007} model is
suitable to study the impacts of the cluster formation on the
isospin sensitive observables, such as n/p ratios for emitted
nucleons and isospin diffusion for central collision and peripheral
collisions at $E/A=50MeV$ for $Sn+Sn$. Fig.1 shows the contour plots
of multiplicity of fragments with charge $Z$ as a function of its
scaled rapidity $y/y_{beam}^{c.m.}$ at $E/A=50MeV$ calculated with
soft symmetry energy ($\gamma_{i}=0.5$) for different impact
parameters, $b=2, 6, 8fm$, respectively.
\begin{figure}[th]
\centerline{\psfig{file=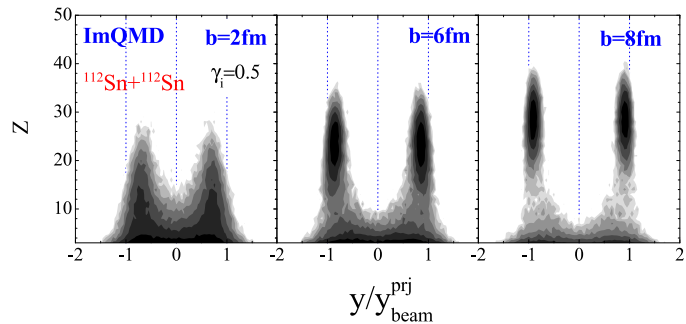,width=3.5cm,angle=270}}
 \vspace*{40pt}
\caption{The contour plots of multiplicity of fragments with charge
$Z$ as a function of its scaled rapidity $y/y_{beam}^{c.m.}$ for
$^{112}Sn+^{112}Sn$ at $E/A=50MeV$ for $b=2,6,8fm$ with
$\gamma_{i}=0.5$ cases.}
\end{figure}
In general, one can observe two peaks on the rapidity distribution
for the charge $Z$ of fragments from central collisions to
peripheral collisions, and the intermediate mass fragments can be
formed from middle rapidity to projectile/target rapidity for
central collisions and peripheral collisions. In detail, the maximum
charge of fragments at mid-rapidity decrease from $Z\sim15$ to
$Z\sim8$ with the impact parameter $b$ increasing, and same for its
yield. Furthermore, the heaviest fragments have lost about $35\%$ of
their initial velocity for central collisions and about $25\%$
($10\%$) of their initial velocity for $b=6fm$ ($b=8fm$). The
quantities of velocity loss for heaviest fragments depend on the
decelerating effects from the effective n-n interactions and
nucleon-nucleon collision in the participant region, and the
multiplicity of fragments formed below the normal density in HICs.
Through above discussion, one can see that the clusters formation
plays an important role at intermediate energy HICs from central to
peripheral collisions and the effects of cluster emission on the
isospin sensitive observables could not be ignored.
\\
\\
In order to study the density dependence of symmetry energy and the
influence of cluster emission on the isospin sensitive observables,
the density dependence of symmetry energy in nuclear matter:
\begin{equation}
E_{sym}(\rho)=\frac{1}{3}\frac{\hbar^2}{2m}\rho_{0}^{2/3}(\frac{3\pi^2}{2}\frac{\rho}{\rho_{0}})^{2/3}
+\frac{C_{s}}{2}(\frac{\rho}{\rho_{0}})^{\gamma_{i}}
\end{equation}
is introduced. $C_{s}$ is the symmetry potential strength
parameters, and $\gamma_{i}$ give the density dependence of symmetry
energy. The different density dependence of symmetry energy can be
realized by varying $\gamma_{i}$ in the ImQMD model.

\section{The n/p ratios, isospin diffusion and the cluster emission effects}

The neutron to proton ratio
$R_{n/p}=\frac{dM_{n}(A)}{dE_{cm}}/\frac{dM_{p}(A)}{dE_{cm}}$ of
pre-equilibrium emitted neutron over proton spectra was considered
as a sensitive observable to the density dependence of symmetry
energy\cite{BALi1997}, because it has a straightforward link to the
symmetry energy. In order to reduce the sensitivity to uncertainties
in the neutron detection efficiencies and sensitivity to relative
uncertainties in energy calibrations of neutrons and protons, the
double ratio
\begin{equation}
DR(n/p)=R_{n/p}(A)/R_{n/p}(B)=\frac{dM_{n}(A)/dE_{cm}}{dM_{p}(A)/dE_{cm}}/\frac{dM_{n}(B)/dE_{cm}}{dM_{p}(B)/dE_{cm}}
\end{equation}
had been measured by Famiano and compared with the transport model
prediction\cite{Fami2006,BALi1997}.

We have performed calculations of collisions at an impact parameter
of $b=2 fm$ at an incident energy of $50 MeV$ per nucleon for two
systems: $A=^{124}Sn+^{124}Sn$ and
$B=^{112}Sn+^{112}Sn$\cite{Zhang2008} with ImQMD05 to study the
$DR(n/p)$ ratio for emitted nucleons. The shaded regions in the left
panel of Fig.2 (a) show the range, determined by uncertainties in
the simulations, of predicted double ratios $DR(n/p)=R_{n/p}(A)/
R_{n/p}(B)$ of the nucleons emitted between $70^{\circ}$  and
$110^{\circ}$  in the center of mass frame as a function of the
center of mass nucleon energy, for $\gamma_{i}=0.5$ and $2$. The
double ratios $DR(n/p)$ are higher for the EOS with the weaker
symmetry energy density dependence $\gamma_{i}=0.5$ than that for
$\gamma_{i}=2.0$ because the nucleons mainly emit from the lower
density region at intermediate energy HICs. Compare to the data on
$DR(n/p)$ for emitted nucleons(solid stars), the general trend of
data $DR(n/p)$ are qualitatively reproduced and the data seem to be
closer to the calculation employing the EOS with $\gamma_{i} =0.5$.
The fig.2 (b) show the coalescence-invariant double ratio. The
coalescence-invariant double ratios are constructed by including all
neutrons and protons emitted at a given velocity, regardless of
whether they are emitted free or within a cluster. The data are
shown as open stars and the calculation results are shown as shaded
region in the right panel of figure 2(b). Here, the measurements and
simulations results illustrate that the fragments with $Z\ge2$
mainly contribute to the low energy spectra and do not affect the
high-energy $DR(n/p)$ data very much.

\begin{figure}[th]
\centerline{\psfig{file=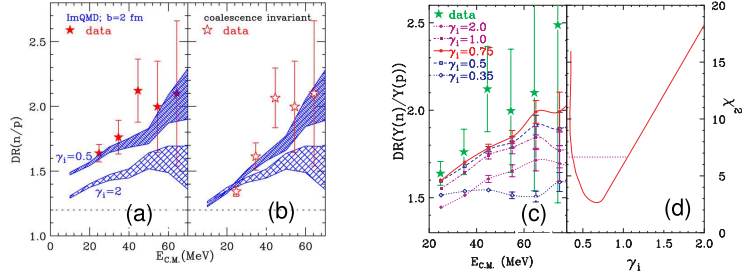,width=3.0cm,angle=270}}
\vspace*{40pt} \caption{(a) DR(n/p) ratios for emitted free nucleons
and (b) coalescent-invariant $DR(n/p)$ from the ImQMD simulations
are plotted as shadow region. Data (star) from
NSCL\protect\cite{Zhang2008}.(c) $DR(n/p)$ ratios for different
$\gamma_{i}$ for central collisions, Data (star) from NSCL. (d)
$\chi^2$ analysis on the $DR(n/p)$ ratios\protect\cite{Tsang2009}.}
\end{figure}
In order to constrain the range of $\gamma_{i}$ from the $DR(n/p)$
data that had been published, a series calculations for two systems,
$A=^{124}Sn+^{124}Sn$ and $B=^{112}Sn+^{112}Sn$, have been performed
by varying $\gamma_{i}=0.35, 0.5, 0.75, 1.0$ and $2.0$(line with
symbols)\cite{Tsang2009}. As shown in fig.2(c), the computation
uncertainties are statistical. It is known that emitted nucleons
mainly from the subnormal densities at this energies. So, the n/p
ratios of emitted nucleons are associated with the values of
symmetry energy at subnormal density. Therefor, the $DR(n/p)$ ratio
should increase with decreasing $\gamma_{i}$. However, in the limit
of very small $\gamma_{i}\ll0.35$, the system completely
disintegrates and the $DR(n/p)$ ratio decrease and approaches the
limit of reaction system, $(N/Z)_{124}/(N/Z)_{112}=1.2$. As a
consequence of these two competing effects, the double ratio values
peak around the $\gamma_{i}=0.7$. Despite the large experiment
uncertainties for higher energy data, those comparisons definitely
rule out very soft ($\gamma_{i}=0.35$) and very stiff
($\gamma_{i}=2.0$) density dependence of symmetry energy. Fig.2(d)
shows the dependence on $\gamma_{i}$ of the $total$ $\chi^2$
computed from the difference between predicted and measured double
ratios. Within a $2\sigma$ uncertainty, parameters of $\gamma_{i}$
fall in the range of $0.4\le\gamma_{i}\le1.05$ for the
$C_{s}=35.2MeV$.
\\
\\
It has been proved that the isospin diffusion ability depends on the
strength of the symmetry energy in
HICs\cite{Tsang2004,LWChen2005,LShi2003}. In order to quantify the
isospin diffusion degree in heavy ion collisions, the isospin
transport ratios $R_{i}$
\begin{equation}
R_{i}=(2X-X_{A+A}-X_{B+B})/(X_{A+A}-X_{B+B})
\end{equation}
has been introduced\cite{Tsang2004}. The subscript $A$ and $B$
represent the neutron rich and neutron-poor nuclei, and the
$A=^{124}Sn$, $B=^{112}Sn$ in this work. The value of $R_{i}$ is
obtained through $3$ reaction systems at least, $A+A$, $B+B$ and
$A+B(or B+A)$. Where $X$ is the isospin tracer from the isospin
asymmetry nuclear reaction system $A+B(or B+A)$. The non-isospin
diffusion effects are minimized by scaling the isospin observables
with the same observables from the symmetric collisions of the
neutron-rich $A+A$ and neutron-deficient $B+B$ systems using the
isospin transport ratio defined as above. Adopting the Eq.3
definition on $R_{i}$, one can expects $R_{i}=\pm 1$ in absence of
isospin diffusion. In the opposite, $R_{i}\approx 0$ if isospin
equilibrium is achieved. Eq.3 also dictates that different
observable, $X$,  can give the same results if they are linearly
related\cite{Tsang2004,TXLiu2007}. In experiment, the $X$ is taken
as the isoscaling parameter $X_{\alpha}=\alpha$ and the yield ratios
of $A=7$ mirror nuclei, $X_{7}=ln(Y(^{7}Li)/Y(^{7}Be))$ for
peripheral HICs\cite{Tsang2004,TXLiu2007}. In order to analyze the
isospin diffusion data with transport model, the observable
$X=\delta$, the isospin asymmetry constructed from the fragments and
free nucleons at the relevant rapidities is adopted in ImQMD. It has
been confirmed theoretically and experimentally that there is a
linear relationship between $X_{exp}=X_{7},X_{\alpha}$ and
$X_{th}=\delta$, and the relationship
$R(X=\alpha)=R_{7}(X=X_{7})=R(\delta)$
holds\cite{Tsang2001,Botvi2002,AOno2003,Tsang2009}.

The left panel in Fig.3 show the ImQMD predictions on the
$R(X=\delta)$ (lines) near the projectile rapidity as a function of
impact parameters $b$ for different $\gamma_{i}=0.35, 0.5, 0.75, 1$
and $2$\cite{Tsang2009}. Faster equilibration occurs for smaller
$\gamma_{i}$ values which correspond to a larger symmetry energy at
subnormal densities. Thus, we see a monotonic decrease of the
absolute values of $R_{i}(\delta)$ with decreasing $\gamma_{i}$.
Experimental data on $R_{i}(\alpha)$\cite{Tsang2004}, is plotted as
shaded regions in the left panel of Fig.3. Performing the $\chi^{2}$
analysis between the $\mathrm{ImQMD}$ predictions and the
experimental data on $R_{i}(\alpha)$, the range of the symmetry
potential parameter $0.45\le\gamma_{i}\le1.0$ are obtained within
$2\sigma$. Furthermore, the ImQMD predictions on the $R(X=\delta)$
(lines) as a function of rapidity are calculated and shown in the
middle panel of fig.3 as lines, the star symbols represent measured
values of $R_{7}$ obtained from the yield ratios of $^{7}Li$ and
$^{7}Be$ at $b=6fm$\cite{TXLiu2007}. It can be seen that the
calculation of the shapes and magnitude of the rapidity dependence
of the isospin transport ratios reproduces the trends accurately.
The corresponding $\chi^{2}$ analysis on $R_{7}$ with calculations
at $b=6$ and $7fm$ favors the region $0.45\le\gamma_{i}\le0.95$
within $2\sigma$ uncertainties.
\begin{figure}[th]

\centerline{\psfig{file=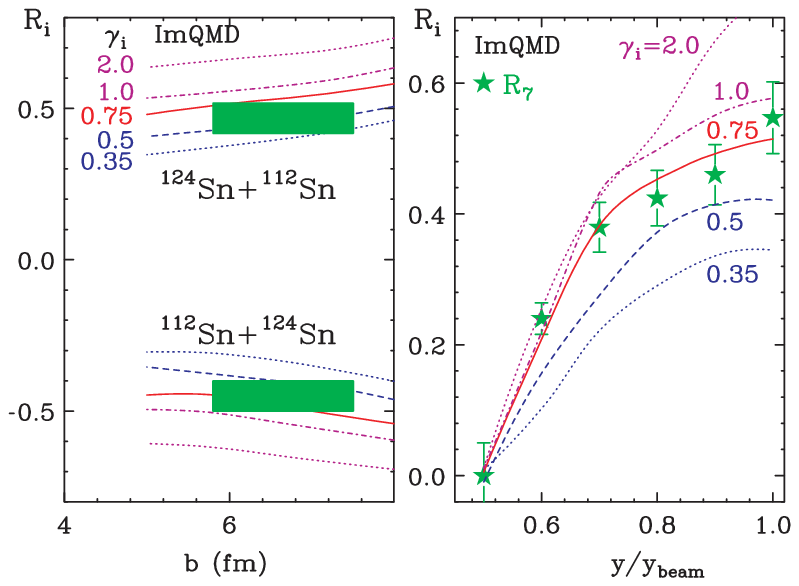,width=1cm,angle=0,totalheight=5.05cm}
\psfig{file=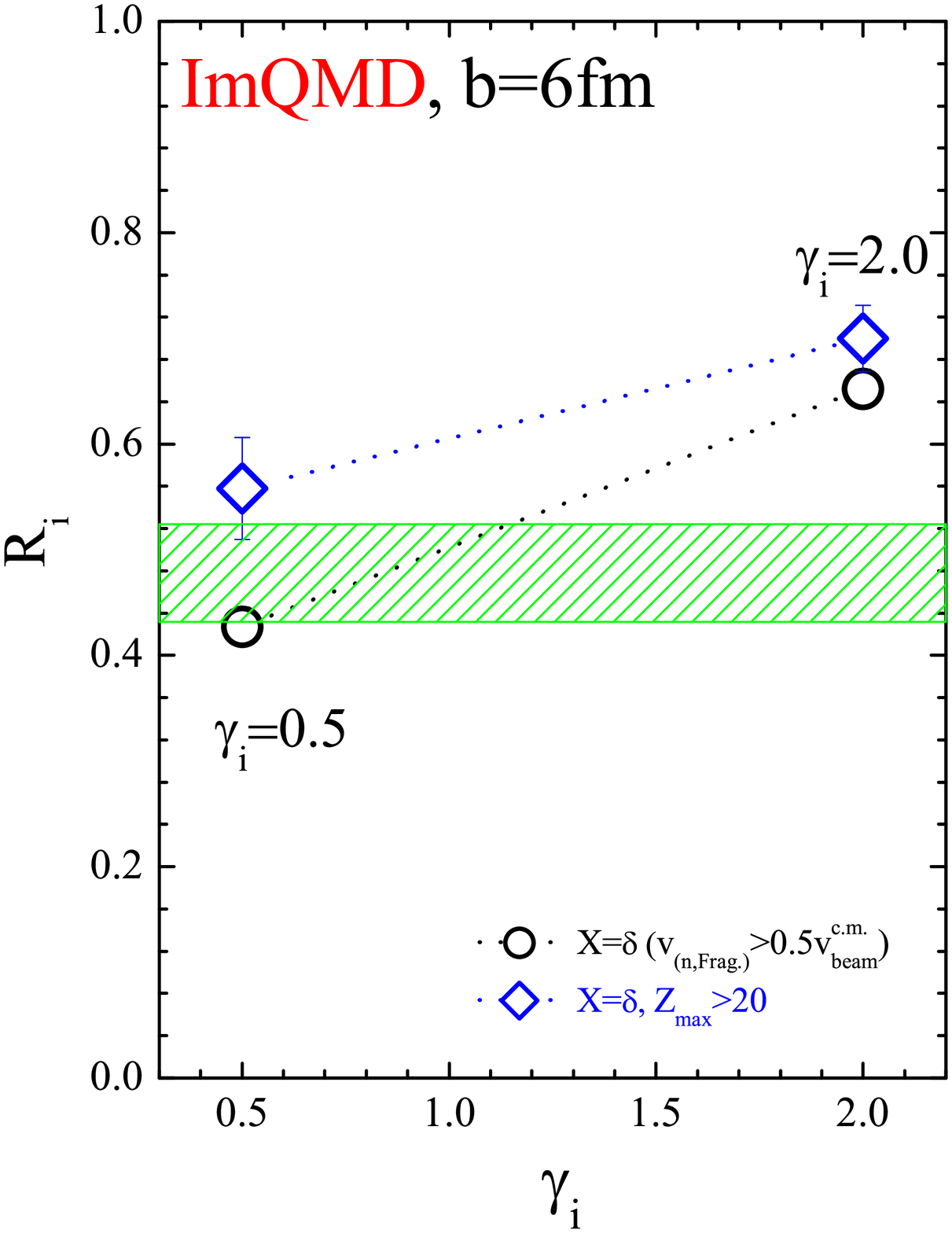,width=1cm,angle=0,totalheight=4.0cm}}

\vspace*{14pt} \caption{Left panel: The calculated results on
isospin transport ratios from the ImQMD model(lines) as a function
of impact parameters for different values of $\gamma_{i}$ and the
experiment data $R_{i}(\alpha)$ (shaded regions). Middle panel:
ImQMD calculations on the isospin transport ratios as a function of
rapidity for different $\gamma_{i}$ at $b=6fm$ and the experiment
data on $R_{7}$ as a function of rapidity (star
symbols)\protect\cite{Tsang2009}.Right panel: Isospin transport
ratios obtained from different isospin tracer, $X=\delta(N,Frags))$
(open circles) and $X=\delta(Z_{max}>20)$ (open diamonds) in
projectile region }
\end{figure}

In order to see the cluster effects on the isospin transport ratio
$R_{i}$, a tracer defined by the isospin asymmetry of the heaviest
fragments with $Z\ge20$ in projectile region is adopted in the
transport model simulation for peripheral HICs ($b=6fm$). The right
panel in fig.3 show the $R_{i}$ calculated with
$X=\delta(Z_{max}>20)$
 (open diamonds) and open circles represent the results from the
 $X=\delta$ near the projectile rapidity as previous discussed. Independent of the isospin tracer we adopted, $R_{i}$ obtained with
soft symmetry case ($\gamma_{i}=0.5$) is smaller than those obtained
with stiff symmetry potential case ($\gamma_{i}=2.0$). Another side,
the values of $R_{i}$ obtained with the new tracer
$X=\delta(Z_{max}>20)$ are larger than that with $X=\delta$
constructed from emitted nucleons and fragments near the projectile
rapidity. It is caused by the cluster emission effects and the
dynamical properties of the isospin diffusions. In HICs process, the
neutrons (protons) of the projectile or target mainly transfer to
the lower density neck region and then break up into IMFs and
nucleons, few nucleons from target residues diffuse to the
projectile residues or vice versa due to the short diffusion time
scale at $E/A=50MeV$. It leads the larger values of $R_{i}$ for the
new tracer $X=\delta(Z_{max}>20)$. So, study the $R_{i}$ with these
different isospin tracer will help us to understand the dynamic
effects on isospin diffusion in HICs and give a new constraint on
the density dependence of symmetry energy.

\section{Summary and outlook}
In summary, the reaction mechanism of the central and peripheral
collisions are investigated for $^{112,124}Sn+^{112,124}Sn$ at
$E/A=50MeV$ for different impact parameters. The results show that
the reactions systems do not reach the global thermal equilibrium
even for $Sn+Sn$ central collisions at $E/A=50MeV$ and the effects
of cluster emission are important for both central and peripheral
collisions. Three observables, double $n/p$ ratios, isospin
diffusion and the rapidity distribution the ratio $R_{7}$ are
analyzed with the Improved Quantum Molecular Dynamics model. The
results show that these three observables are sensitive to the
density dependence of symmetry energy. By comparing calculation
results to the data, the consistent constraints on the density
dependence of symmetry energy can be obtained. Furthermore, the
analysis of the coalescence-invariant $DR(n/p)$ ratio and the
isospin transport ratios $R_{i}(X=\delta(Z_{max}>20))$ show that the
influence of the cluster emission on the double n/p ratios and the
isospin transport ratio are important. However, the large
uncertainties on the data implies that the improvements in the
precision of data should be done in the near future, and accurate
constraints on the density dependence of symmetry energy should be
complemented from many aspects, such as giant dipole resonances,
pygmy dipole resonance, and mass data and some new observables from
HICs and the observed properties of neutron star.


This work has been supported by the Chinese National Science
Foundation under Grants 10675172, 10175093, 10235030 and the U.S.
National Science Foundation under Grants PHY-0216783, 0456903,
060007, 0800026 and the High Performance Computing Center(HPCC) at
Michigan State University.

\end{document}